\documentclass[aps,prl,twocolumn,showpacs,preprintnumbers,amsmath,amssymb,
superscriptaddress]{revtex4}

\usepackage{graphicx}
\usepackage{dcolumn}
\usepackage{bm}
\begin{document}
\title{Frustration in The Coupled Rattler System KOs$_2$O$_6$}
\author{J. Kune\v{s}}
\affiliation{Department of Physics, University of California, 
Davis CA 95616, USA}
\email{jkunes@physics.ucdavis.edu}
\affiliation{Institute of Physics,
Academy of Sciences of the Czech Republic, Cukrovarnick\'a 10,
162 53 Praha 6, Czech Republic}
\author{W.\,E. Pickett}
\affiliation{Department of Physics, University of California, 
Davis CA 95616, USA}
\date{\today}


\maketitle

{\bf
The phenomenon of frustration, which gives rise to many fascinating phenomena, 
is conventionally
associated with the topology of non-bipartite lattices, where 
nearest-neighbor (nn) interactions and global connectivity compete
in the lowering of energy. 
The issue of rattling atoms in spacious lattice sites 
is a separate
occurrence that can also lead to a high density of low energy states (unusual low temperature
thermodynamics) and to practical applications such as in improved thermoelectric materials.
In this letter we address a unique situation where both phenomena
arise: a four-fold single-site instability leads to rattling of cations
on a diamond structure sublattice where nn interactions frustrate simple
ordering of the displacements. The system deals with this coupling of
rattling+frustration by commensurate ordering.  Such a disorder-order
transition may account for the second phase transition seen in KOs$_2$O$_6$ 
within the superconducting state, and the unusual low-energy dynamics
and associated electron-phonon coupling can account for the qualitative
differences in physical properties of KOs$_2$O$_6$ compared to RbOs$_2$O$_6$ and 
CsOs$_2$O$_6$, all of which have essentially identical average crystal and electronic structures. 
}

The pyrochlore-lattice-based structure with a potential to support magnetic 
frustration has attracted attention to AOs$_2$O$_6$ (A=K, Rb,Cs) group. 
Unexpectedly large variation of the superconducting $T_c$ throughout the group 
(from 3.3 K in CsOs$_2$O$_6$ to 9.7 K in KOs$_2$O$_6$)
\cite{roso,coso,mur05} together with reports of anomalous nuclear spin relaxation \cite{nmr} and 
indications of anisotropic order parameter \cite{msr} in KOs$_2$O$_6$ 
pointed to a possibility of unconventional pairing and fueled the
early experimental interest. While the issue of superconductivity remains
controversial in the light of recent pressure experiments \cite{mur05},
unusual transport and thermodynamic properties were found in the normal
state of KOs$_2$O$_6$ in sharp contrast to the standard metallic 
behavior of RbOs$_2$O$_6$ and CsOs$_2$O$_6$ \cite{kha05,sch05,nmrrb}.

Uniquely to KOs$_2$O$_6$ within this class,
the normal-state conductivity exhibits a non-Fermi-liquid behavior 
characterized by a concave temperature dependency down to low temperatures \cite{cvkoso,bruh06}. 
The low temperature linear specific heat coefficient is estimated to be
substantially larger than in RbOs$_2$O$_6$ and CsOs$_2$O$_6$ \cite{cvkoso}. 
Recently an intriguing $\lambda$-shaped peak in the
specific heat was observed in good quality KOs$_2$O$_6$ single-crystals indicative 
of a phase transition at T$_p$ = 7 K \cite{cvkoso}, within the superconducting state. 
This observation was recently confirmed \cite{bruh06}.
Notably, the peak position and shape do not change
even when the superconductivity is suppressed below 7 K by the external field. 
Insensitivity to such a profound change of the electronic state indicates that the peak is 
rooted in the lattice dynamics rather than intrinsic electronic degrees of freedom.

Electronic structure investigations \cite{saniz,kunes} have revealed a considerable 
bandwidth of the Os-$5d$-$t_{2g}$ 12-band complex of about 3 eV which
does not support the idea of local moment formation on the Os sites nor any emergence of frustration due to
the pyrochlore topology of the Os sublattice, made of a three-dimensional network of
vertex-sharing tetrahedra. Instead we find that a significant frustration,
not magnetic but structural,
takes place on the diamond sublattice occupied by K ions. We have shown previously \cite{kunes} that the symmetric (A$_g$) 
potassium phonon mode is unstable
and that the energy can be lowered by several meV/atom(K) through rather large displacements
of the K ions. Here we construct, based on first principles calculations, the effective potential describing
fourfold symmetric displacements of K ions off their ideal diamond-lattice sites, with nn
coupling leading to a highly frustrated system of displacements. Dynamical simulations for finite clusters
reveal a classical ground state with complex pattern of displacements.
\begin{figure}
\includegraphics[width=0.7\columnwidth]{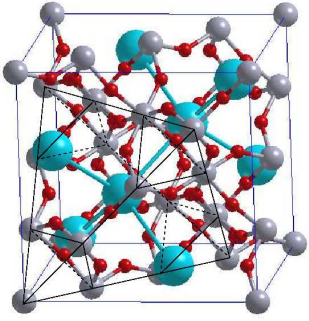}
\caption{\label{fig:structure} {\bf The AOs$_2$O$_6$ lattice.} The atomic species are maker with different colors:
Os (gray), O (red) and alkali metal (blue). The pyrochlore sublattice
of Os atoms is highlighted. Notice the alkali sublattice with diamond structure.}
\end{figure}

\begin{figure}
\includegraphics[height=\columnwidth, angle=270]{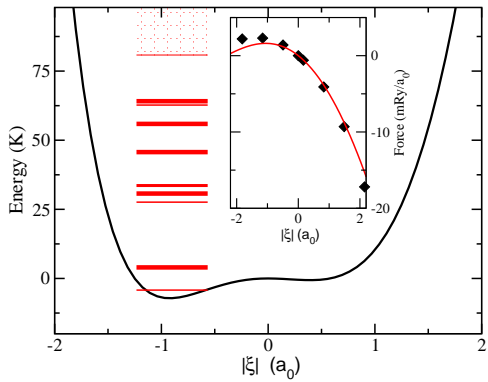}
\caption{\label{fig:onsite} {\bf The on-site potential.}  
The on-site potential along the nn bond direction (the nn site is in the direction
of positive x-axis). The vertical line denote the eigenenergies on the same scale,
the thickness represents degeneracy ranging from 1 to 3.
In the inset force acting on undisplaced ion as a function of the uniform displacement
of its neighbors is shown. The symbol represent the {\it ab initio} data, 
the full line is the fit with linear pair force.}
\end{figure}

In Fig. \ref{fig:structure} we show the AOs$_2$O$_6$ lattice which consists of Os on a pyrochlore sublattice, 
having one O atom bridging each Os nn pair. The cavities in the Os-O network are filled with alkali ions, 
which themselves form a diamond lattice, composed of two $fcc$ sublattices. Using a full-potential 
linearized augmented-plane-waves code Wien2k \cite{wien} we have performed a series of calculations 
in which K ions move along the (111) direction: (i) the two $fcc$ sublattices are displaced in opposite direction (symmetric A$_g$ 
mode), (ii)  same as (i) with the O positions allowed to relax, (iii) only one $fcc$ sublattice is displaced.
Comparing results (i) and (ii) reveals a non-negligible O relaxation only for large K displacement 
$\boldsymbol{\xi_i}$ (the energy vs 
displacement curve has slightly less steep walls when O ions are allowed to relax). Since the
O relaxation effect is minor we consider the Os-O network to be rigid for the 
following discussion.
The inter-ionic distances together with the geometry of the Os-O network with   
spacious channels along the nn K-K bonds suggest nn coupling to dominate over
longer range interaction. The effective Hamiltonian becomes
\begin{equation}
\label{eq:ham}
\begin{split}
\hat{H}=\sum_i \bigl[\frac{p_i^2}{2M}+P_e(\xi_i)+P_o(\xi_i)\mathcal{Y}_{32}(\hat{\boldsymbol{\xi_i}})\bigr]+\\
+\sum_{i>j}W_{ij}(\boldsymbol{\xi_i},\boldsymbol{\xi_j}),
\end{split}
\end{equation}
where the first term is the on-site Hamiltonian and second describes the nn coupling (interaction). The on-site potential, which 
captures the essential tetrahedral local symmetry, consists of a spherical and the next non-zero term 
in spherical harmonic expansion ($\mathcal{Y}_{lm}$), 
while the radial dependency is described by even and odd 6th-order polynomials $P_e(\xi_i)$ and $P_o(\xi_i)$
obtained by fitting the {\it ab initio} data from type (iii) calculations (Fig. \ref{fig:onsite}). 

We have solved the quantum-mechanical single site problem
by numerical integration on real space grid (details of the calculation can be found in Ref. \onlinecite{sces}).
The low energy spectrum up to 80 K (containing 20 states)
is characterized by a singlet-triplet split ground state (8 K splitting) separated by a 
gap of about 25 K from excited states. This essential difference from the harmonic
potential with singlet ground state is reflected also by a Schottkyesque anomaly in the
single-site specific heat.\cite{sces} The sharpness 
of the observed peak, however,
rules out the Schottky anomaly scenario, pointing instead to a collective transformation
involving inter-site coupling. A particularly useful way of looking at the quasi-degenerate
quadruplet ground state, is in terms of four symmetry related local orbitals centered
in the local minima. A weak coupling of about -2 K for each pair of orbitals accounts
for the singlet-triplet splitting.

{\it Ab initio} calculations \cite{kunes} revealed much less anharmonicity in the
Rb and especially Cs potentials, which can treated as a perturbation and neglected
for the present purposes.  Within this approximation
Rb (Cs) dynamics is described by 3D oscillator with a frequency of 44 K (61 K). 
Recent analysis of specific heat data by
Br\"uhwiler {\it et al.} \cite{bruh06} led to a frequency of 60 K
for RbOs$_2$O$_6$, which we find a reasonable agreement given the approximations.
Localized modes were previously observed in specific heat of RbOs$_2$O$_6$ and CsOs$_2$O$_6$
by Hiroi {\it et al.}\cite{cvaoso} 
The harmonic oscillator root mean square displacement is
\begin{equation}
\Delta=\sqrt{\langle {\xi_i^{\alpha}}^2 \rangle} =\frac{1}{\sqrt{2M\omega}},
     ~~\alpha = x, y, z,
\end{equation}
and we obtain $\Delta_{Rb}=0.15a_0$ (Bohr radius) and $\Delta_{Cs}=0.1a_0$ for the mean square displacement 
at zero temperature, which we will use below in estimation of strength of the inter-site coupling.

The interaction can be obtained by following the force acting on a fixed ion when its neighbors are
uniformly displaced.
The simplest form of central pair force that describes reasonably 
well the {\it ab initio} data is
$F(\mathbf r)=A\tfrac{\mathbf r}{r}+B{\mathbf r}$, corresponding to a pair potential 
$V(\mathbf{r_1},\mathbf{r_2})=A|\mathbf{r_{12}}|+\tfrac{B}{2}|\mathbf{r_{12}}|^2$ ($\mathbf{r_{j}}$ are ion coordinates). 
In Fig. \ref{fig:onsite}(inset) we show the first principles force together with the model fit,
the values A=-88 mRy/a$_0$ and B=7.9 mRy/a$_0^2$ are essentially the same for all three oxides.
As expected from its electrostatic origin the pair force is repulsive for admissible
values of r. Using the electrostatic force $-\tfrac{1}{r^2}$ instead of an {\it ad hoc} Taylor
expansion, the force in Fig. \ref{fig:onsite} would not crossover to the positive values, but
only reach zero for zero displacement. The observed behavior is qualitatively consistent with
faster decay of the interaction due to screening.
The interaction $W_{ij}(\boldsymbol{\xi_i}, \boldsymbol{\xi_j})$ between ions at sites
$\mathbf{R_i}$ and $\mathbf{R_j}$ is obtained after 
subtraction of contributions accounted for in the on-site potential:
\begin{equation}
\label{eq:interaction}
\begin{split}
W_{ij}(\boldsymbol{\xi_i}, \boldsymbol{\xi_j})=&V(\mathbf{R_i}+\boldsymbol{\xi_i},\mathbf{R_j}+\boldsymbol{\xi_j})-
V(\mathbf{R_i},\mathbf{R_j}+\boldsymbol{\xi_j})\\-&V(\mathbf{R_i}+\boldsymbol{\xi_i},\mathbf{R_j})+V(\mathbf{R_i},\mathbf{R_j}).
\end{split}
\end{equation}
The result is a directional (non-central) potential, whose dipolar form becomes clear 
in the small displacement limit: 
\begin{equation}
\label{eq:dip}
\Tilde{W}_{ij}(\boldsymbol{\xi_i},\boldsymbol{\xi_j})\approx A\frac{(\mathbf{R_{ij}}\cdot\boldsymbol{\xi_i})(
\mathbf{R_{ij}}\cdot\boldsymbol{\xi_j})}{R_{ij}^3}-(\frac{A}{R_{ij}}+B)\boldsymbol{\xi_i}\cdot\boldsymbol{\xi_j}.
\end{equation}

We start the discussion of nn coupling in the simpler quasiharmonic case 
(RbOs$_2$O$_6$ and CsOs$_2$O$_6$).
Small mean displacements justify the use of a dipolar approximation (\ref{eq:dip})
to estimate the interaction energy. Using the Schwarz inequality
$|\langle \xi_i \cdot \xi_j \rangle|^2\leq \langle \xi_i^2 \rangle\langle \xi_j^2 \rangle$
the interaction energy per site for a two-site problem has the following upper bound
\begin{equation}
|\langle \Tilde{W} \rangle|\leq \bigl(\frac{|B|}{2}+|\frac{A}{R}+B|\bigr) \Delta^2,
\end{equation}
yielding 23 K and 10 K for Rb and Cs respectively.  
Going one step further and considering the problem of four sites connected to their common neighbor
the upper bound is reduced for purely geometrical reasons to 16 K and 7 K respectively.
Smallness of the interaction energy in comparison to 
the Einstein frequencies and the singlet character of the on-site groundstate
lead to conclusion that Rb and Cs dynamics is essentially local.

KOs$_2$O$_6$ presents the {\it opposite limit} of extreme anharmonicity, 
leading to
quasi-degeneracy of the ground state and large spatial fluctuations,
together with important inter-site coupling. In the following
we evaluate the matrix elements of $W$ in the basis of products $|\alpha \rangle$
of the on-site eigenstates.
The basis of 
local orbitals for the ground state quadruplet has an advantage
of keeping the Hamiltonian in quasi-diagonal form by maximizing the diagonal terms
$\langle \alpha \beta|W|\beta \alpha \rangle$ and minimizing the leading off-diagonal contributions
$\langle \alpha \beta|W|\gamma \alpha \rangle$. Moreover there is a natural one-to-one mapping between
the local orbitals and nn bonds, namely we denote an orbital and a bond with the same
index if the orbital represents displacement in the direction of the bond.
Due to the symmetry there are only four independent diagonal matrix elements, which can 
be expressed in the form (in kelvins):
\begin{equation}
\label{eq:bond}
\begin{split}
\langle \alpha \beta |W_R| \beta \alpha \rangle=-324\delta_{\alpha\beta}+742\delta_{\alpha R}
\delta_{\beta R}\\
-301(\delta_{\alpha R}+\delta_{\beta R})+147,
\end{split}
\end{equation}
where both the bond index R and the orbital indices $\alpha$, $\beta$ run from 1 to 4.
These numbers can be understood in terms of the approximate formula (\ref{eq:dip}),
taking into account the inversion symmetry about the bond center. The relevant
off-diagonal terms
yield about 15 K
(additional 2 K comes from the on-site Hamiltonian).
The off-diagonal terms also provide coupling to products including excited states with 
the largest ones being about $1/3$ of the corresponding energy difference, providing thus
small but non-negligible quantum mechanical coupling. 

{\it Origin of frustration.}
Building a lattice model from bonds (\ref{eq:bond}) helps to understand frustrated nature  
of the present system.
Although it is not justifiable
to neglect the excited states completely, they
will only renormalize the parameters without
changing the form of the four-state form of the Hamiltonian in the low energy 
sector. Building a lattice Hamiltonian from the bonds (\ref{eq:bond}), using the fact
that the third term yields a constant when summed over the bonds, we get 
an expression
\begin{equation}
\label{eq:potts}
H=\sum_{ij}(a\delta^{ij}_{\alpha\beta}+b^{\infty}\delta^{i}_{\alpha R(ij)}\delta^{j}_{\beta R(ij)})+H_{on-site}.
\end{equation}
The first term of (\ref{eq:potts}) is the classical Potts Hamiltonian \cite{potts}, 
$\delta^{ij}_{\alpha\beta}$ yields 1 when the neighboring sites $i$, $j$ are occupied by the same state and zero otherwise.
In the second term $R(ij)$ is an index of the bond between sites $i$ and $j$, 
$\delta^{i}_{\alpha R(ij)}$ yields 1 if orbital on site $i$ corresponds to displacement in the
direction of the bond $ij$ and zero otherwise.
The bare values of parameters $a$ and $b$ are -162 K and 371 K respectively.
Since the second term describes states whose energy is above the already neglected excited states
it is consistent to rule these states out by putting $b$ equal to $+\infty$. 
The second term thus becomes a constraint on admissible configurations and introduces 
frustration into the system.
The leading quantum mechanical correction $H_{on-site}$, bare value of which is an order of magnitude
smaller than $a$, is provided by tunneling between the local orbitals. 
Filling the lattice such that we minimize the contribution of an arbitrary first site (only 3 bonds can yield
$a$ due to the constraint) one can readily see that an arrangement with the same energy cannot be
placed on the neighboring sites. Unlike in the case of geometrical frustration of nn antiferromagnets
no odd-length loops are necessary to produce frustration. In fact the above mechanism would apply even to Bethe lattice
with no loops at all. While we cannot make conclusions about the degeneracy of the groundstate,
the frustrating constraint is expected to reduce the transition temperature below the
energy scale defined by parameter $a$.

{\it Dynamical simulations.} While the effective Hamiltonian (\ref{eq:potts}) can be useful for investigating
general features of the phase transition and is well suited for analytical approach, 
in the rest of this paper we pursue a separate, purely numerical 
approach to probe aspects of the ordering that we anticipate at T$_p$. 
Addressing this question in full generality is very difficult. 
Insight can be gained by minimizing the potential
energy, i.e. pursuing the classical (large mass M) limit, for finite clusters with
periodic boundary conditions. This is still a formidable computational task due to 
a large number of local minima. To approach and possibly reach the global minimum we have used 
a damped molecular dynamics combined with simulated annealing. 
In particular we have integrated the classical equation of motion
\begin{equation}
\label{eq:md}
M\frac{d^2\boldsymbol{\xi}}{dt^2}=\mathbf{\mathcal{F}}(\boldsymbol{\xi})-\beta(T)\frac{d\boldsymbol{\xi}}{dt}
+\mathbf{\mathcal{G}}(T)
\end{equation} 
where $\mathbf{\mathcal{F}}$ is the actual force,
the $\beta(T)\propto \sqrt{T}$ is a friction parameter and $\mathbf{\mathcal{G}}(T)$ is a Gaussian random vector
with half-width proportional to $T$.
The effective temperature $T$ was successively reduced, $T_i=\epsilon T_{i-1} 
(\epsilon < 1)$, until
minimum was reached.

\begin{figure}
\includegraphics[width=0.8\columnwidth]{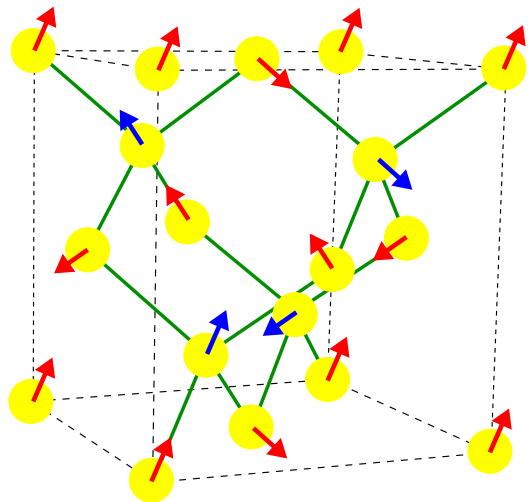}
\caption{\label{fig:222} Minimum energy configuration for $2\times2\times2$ cluster. Different colors
correspond the two fcc sublattices, green lines mark the nn bonds. (Note that
the same displacements for sites on the same edge of the cube are not enforced by the boundary conditions.)}
\end{figure}

The minimum of the $1\times1\times1$ (single primitive cell) cluster 
can be described as parallel displacement of all ions along one of the bond directions
with different displacement values (1.54a$_0$ toward the nn site and 1.24a$_0$ away from nn site) on the two 
sublattices (the global minimum is of course degenerate with respect to the sublattice exchange).
The ordering on a $2\times2\times2$ cluster is characterized by uniform displacements
along different bonds as shown in Fig. \ref{fig:222}. 
The minima for $3\times3\times3$ and larger clusters are difficult to understand
in real space since the displacements are neither uniform nor limited to bond directions.
Nevertheless, common features include a small net displacement
per sublattice (less then 0.1a$_0$) and an average displacement of 2.0a$_0$ per site
with a standard deviation of about 0.3$a_0$. The size of displacements
is likely to be overestimated due to neglect of the kinetic energy, effect of which
can be qualitatively visualized as replacing the point particles with probability density clouds.
Moreover Fourier transform of the displacement vectors $\xi_{\alpha}(\mathbf R_i)$
\begin{equation}
S_{\alpha}(\mathbf q)=\frac{1}{N}\sum_{i}\exp(i\mathbf{q}\cdot\mathbf{R_i})\xi_{\alpha}(\mathbf R_i),
\end{equation}
revealed that there are only a few non-vanishing q-components for each cluster size.
Even with the lowest cooling rate we were not able to obtain the minimum for $5\times5\times5$ cluster 
unambiguously, which strongly suggests that periodicity of 5 unit cells is not commensurate with the ordering
tendencies in the system. The results are summarized in Table {\ref{tab:md}}. 
Fourier transforms are characterized by 2 or 3 dominant components
with (2/3,2/3,0) and (1/2,1/2,1/2) appearing whenever allowed by the cluster size.
Comparison of the minimum energies for different clusters indicates 
that beyond $3\times3\times3$ the energetics becomes very flat while the ordering wavevectors are
sensitive to boundary conditions.
The absolute value of this energy has no
relevance for low energy scale of the ordering transition, but its convergence 
indicates 
that minimum energy is being reached. 

\begin{table}[b]
\caption{\label{tab:md} Minimum potential energy and the dominant 
Fourier components (only one member of $\pm \mathbf q$ pair is shown) for the ground states of clusters of different
size (in brackets). The vectors are in the units of $\tfrac{2\pi}{a}$
where $a$=10.101 $\AA$.}
\begin{tabular*}{\columnwidth}{|l|c|c|}
\hline
cluster size & E (mRy) & largest $|S(q)|$ for \\
\hline \hline
$1\times1\times1$ (2) & -1.99 & \\
$2\times2\times2$ (16) & -17.11 & (1,0,0),(0,1,0),(0,0,1) \\
$3\times3\times3$ (54) & -18.01 & $(-\tfrac{2}{3},0,\tfrac{2}{3}),(\tfrac{2}{3},0,\tfrac{2}{3})$ \\
$4\times4\times4$ (128) & -18.11 & $(\tfrac{1}{2},\tfrac{1}{2},-\tfrac{1}{2}),(-\tfrac{1}{4},\tfrac{1}{4},\tfrac{3}{4}),
(\tfrac{1}{4},\tfrac{3}{4},\tfrac{1}{4})$ \\
$6\times6\times6$ (432) & -18.20 & $(0,-\tfrac{2}{3},\tfrac{2}{3}),(-\tfrac{1}{2},\tfrac{1}{2},\tfrac{1}{2}),
(\tfrac{1}{2},\tfrac{1}{6},\tfrac{5}{6})$ \\
\hline
\end{tabular*}
\end{table}

Our results provide a picture of potassium dynamics in KOs$_2$O$_6$ governed
by an effective Hamiltonian characterized by an unusually soft and 
broad local potential, which allows for large
excursions of K ions resulting in significant and frustrating nn coupling.
To address the question of ordering tendencies we have used
classical simulations for finite clusters which provide a complicated but distinct
pattern with multiple-q ordering and large displacements. 
Since the purely numerical model is not well suited for addressing
general questions concerning the phase transition and for understanding
the essence of the present physics we have also proposed an analytic model
with only a few parameters. 
This model is formally a three-dimensional ferromagnetic four-state Potts model 
with an additional constraint on possible configurations.
While the unconstrained model is known to exhibit a first order mean-field-like 
phase transition \cite{potts}, the constraint cannot be relieved in a simple
way by the system and is likely to change 
behavior of the model.

Our calculations suggest a natural explanation
for the second peak observed in the specific heat of KOs$_2$O$_6$ \cite{cvkoso} 
as a phase transition
of the potassium sublattice to supercell order. Anomalies of low temperature 
electronic properties
such as non-Fermi-liquid conductivity and large linear specific heat coefficient \cite{cvkoso} 
can be explained as consequence of atomic motion, which does not freeze down to
the ordering transition at 7 K. We point out that large excursions of K ion 
should affect the NMR measurements due to quadrupolar interaction
and might be responsible for observed anomalies \cite{nmr}.
The dynamics of Rb and Cs ions is very different: the local behavior is quite different
due to the larger ionic radii which gives a different energy scale, 
and this distinction in turn negates the  
inter-site interaction, leaving a simple quasiharmonic local mode.
This result of our first principle calculations
fits well with the observed specific heat and conductivity \cite{cvaoso,bruh06}. 

If it proves possible, synthesis of K$_x$Rb$_{1-x}$Os$_2$O$_6$ will provide a means
of introducing `vacancies' into the model Hamiltonian (\ref{eq:potts}).
The combination of rattling and frustrating nn interaction, facilitated
by 'fine tuning' of the potassium ionic radius to the size of osmium-oxygen
cage, provides a novel physical system, which exhibits a phase transition
as low temperature. 

We acknowledge discussions with R.~R.~P. Singh and R. Seshadri, and communication
with Z. Hiroi, M. Br\"uhwiler, and  B. Batlogg. J.K. was supported by DOE grant FG02-04ER 46111 and 
Grant No. A1010214 of Academy of Sciences of the Czech Republic, and
W.E.P was supported by National Science Foundation grant No. DMR-0421810.

\end{document}